\documentclass[11pt]{article}
\usepackage{amsmath, amssymb,amscd, amsthm}
\usepackage[mathscr]{eucal}
\usepackage{graphics}
\usepackage{fullpage}
\usepackage{url}
\usepackage{amsmath,amsfonts,amssymb,amsthm,url}
\usepackage{fullpage}
\usepackage{graphicx,graphpap}

\newcommand{\comment}[1]{}

\newcommand{\Z}{\ensuremath{\mathbb{Z}}}

\newcommand{\G}{\ensuremath{\mathbb{G}}}

\newcommand{\Zp}{\ensuremath{{\Z_p}}}

\newcommand{\CT}{\ensuremath{\mathrm{CT}}}

\newcommand{\SK}{\ensuremath{\mathrm{SK}}}

\newcommand{\PP}{\ensuremath{\mathrm{PP}}}
\newcommand{\MSK}{\ensuremath{\mathrm{MSK}}}

\newcommand{\AlgA}{\ensuremath{{\cal A}}}
\newcommand{\AlgB}{\ensuremath{{\cal B}}}

\newcommand{\GG}{\ensuremath{\mathcal{G}}}

\newcommand{\bit}{\{0,1\}}

\newcommand{\depth}{\mathtt{depth}}
\newcommand{\gates}{\mathrm{Gates}}
\newcommand{\wires}{\mathrm{Wires}}
\newcommand{\outputwire}{\mathrm{outputwire}}
\newcommand{\inputs}{\mathrm{inputs}}
\newcommand{\gatetype}{\mathtt{GateType}}
\newcommand{\AND}{\mathrm{AND}}
\newcommand{\OR}{\mathrm{OR}}

\begin{document}

\title{Attribute-Based Encryption for Circuits \\
from Multilinear Maps}

\author{Amit Sahai \and Brent Waters}

\date{}
\maketitle

\begin{abstract}

In this work, we provide the first construction of Attribute-Based
Encryption (ABE) for general circuits.  Our construction is based on
the existence of multilinear maps.  We prove selective security of
our scheme in the standard model under the natural multilinear
generalization of the BDDH assumption.  Our scheme achieves both
Key-Policy and Ciphertext-Policy variants of ABE.

\end{abstract}

\newpage

\section{Introduction}

In traditional public key encryption a sender will encrypt a message
to a targeted individual recipient using the recipient's public key.
However, in many applications one may want to have a more general
way of expressing who should be able to view encrypted data. Sahai
and Waters~\cite{SW05} introduced the notion of Attribute-Based
Encryption (ABE).  There are two variants of ABE:  Key-Policy ABE
and Ciphertext-Policy ABE~\cite{GPSW06}.  (We will consider both
these variants in this work.) In  a Key-Policy ABE system, a
ciphertext encrypting a message $M$ is associated with an assignment
$x$ of boolean variables. A secret key $\SK$ is issued by an
authority and is associated with a boolean function $f$ chosen from
some class of allowable functions $\cal F$. A user with a secret key
for $f$ can decrypt a ciphertext associated with $x$, if and only if
$f(x)=1$.

\comment{Amit, note I left out some applications e.g. SSW12 since I wanted to avoid too much self-citation
at the beginning}

Since the introduction of ABE there have been advances in multiple
directions. These include new proof techniques to achieve adaptive
security~\cite{LOSTW10,OT10,LW12}, decentralizing trust among
multiple authorities~\cite{C07,CC09,LW11a}, and applications to
outsourcing computation~\cite{PRV12}.

However, the central challenge of expanding the \emph{class} of
allowable boolean functions $\cal F$ has been very resistant to
attack. Viewed in terms of circuit classes, the work of Goyal
\emph{et al}~\cite{GPSW06} achieved the best result until now: their
construction achieved security essentially for circuits in the
complexity class \textbf{$\mathbf{NC^1}$}.  This is the class of
circuits with depth $\log n$, or equivalently, the class of
functions representable by polynomial-size boolean formulas.
Achieving ABE for general circuits is arguably the central open
direction in this area\footnote{We note that if collusions between
secret key holders are bounded by a publicly known
polynomially-bounded number in advance, then even stronger results
are known~\cite{SS10,GVW12}. However, throughout this paper we will
deal only with the original setting of ABE where unbounded
collusions are allowed between adversarial users.}.

\paragraph{Difficulties in achieving Circuit ABE and the Backtracking Attack.}
To understand why achieving ABE for general circuits has remained a
difficult problem, it is instructive to examine the mechanisms of
existing constructions based on bilinear maps. Intuitively, a
bilinear map allows one to decrypt using groups elements as keys (or
key components) as opposed to exponents. By handing out a secret key
that consists of group elements an authority is able to
computationally hide some secrets embedded in that key from the key
holder herself. In contrast, if a secret key consists of exponents
in $\Zp$ for a prime order group $p$, as in say an ElGamal type
system, then the key holder or collusion of key holders can solve
for these secrets using algebra. This computational hiding in
bilinear map based systems allows an authority to personalize keys
to a user and prevent collusion attacks, which are the central
threat.

Using GPSW~\cite{GPSW06} as a canonical example we illustrate some
of the main principles of decryption. In their system, private keys
consists of bilinear group elements for a group of prime order $p$
and are associated with random values $r_y \in \Zp$ for each leaf
node in the boolean formula $f$. A ciphertext encrypted to
descriptor $x$ has randomness $s \in \Zp$. The decryption algorithm
begins by applying a pairing operation to each ``satisfied'' leaf
node and obtains $e(g,g)^{r_y s}$ for each satisfied node $y$. From
this point onward decryption consists solely of finding if there is
a linear combination (in the exponent) of the $r_y$ values that can
lead to computing $e(g,g)^{\alpha s}$ which will be the ``blinding
factor'' hiding the message $M$. (The variable $e(g,g)^\alpha$ is
defined in the public parameters.) The decryption algorithm should
be able to  find such a linear combination only if $f(x)=1$. Of
particular note is that  once the $e(g,g)^{r_y s}$ values are
computed the pairing operation plays no further role in decryption.
Indeed it cannot, since it is intuitively ``used up'' on the initial
step.

Let's now take a closer look at how GPSW structures the private keys
given a boolean formula. Suppose in a boolean formula that there
consisted an $\OR$ gate $T$ that received inputs from gates $A$ and
$B$. Then the authority would associate gate $T$ with a value $r_T$
and gates $A,B$ with values $r_A=r_B=r_T$ to match the $\OR$
functionality. Now suppose that on a certain input assignment $x$
that gate $A$ evaluates to 1, but gate $B$ evaluates to 0. The
decryptor will then learn the ``decryption value'' $e(g,g)^{s r_A}$
for gate $A$ and can interpolate up by simply by noting that
$e(g,g)^{s r_T}=e(g,g)^{s r_A}$. While this structure reflects an OR
gate, it also has a critical side effect. The decryption algorithm
also learns the decryption value $e(g,g)^{s r_B}$ for gate $B$
\emph{even though gate $B$ evaluates to 0} on input $x$. We call
such a discovery a \emph{backtracking attack}.

Note that boolean formulas are circuits with fanout one. If the
fanout is one, then the backtracking attack produces  no ill effect
since an attacker has nowhere else to go with this information that
he has learned. However, suppose we wanted to extend this structure
with circuits of fanout of two or more, and that gate $B$ also fed
into an $\AND$ gate $R$. In this case the backtracking attack would
allow an attacker to act like $B$ was satisfied in the formula even
though it was not. This misrepresentation can then be propagated up
a different path in the circuit due to the larger fanout.
(Interestingly, this form of attack does not involve collusion with
a second user.)

We believe that such backtracking attacks are the principle reason
that the functionality of existing ABE systems has been limited to
circuits of fanout one. Furthermore, we conjecture that since the
pairing operation is used up in the initial step, that there is no
black-box way of realizing general ABE for circuits from bilienar
maps.

\paragraph{Our Results.}
We present a new methodology for constructing Attribute-Based
Encryption systems for circuits of arbitrary fanout. Our method is
described using multilinear maps. Cryptography with multilinear maps
was first postulated by Boneh and Silverberg where they discussed
potential applications such as one round, $n$-way Diffie-Hellman key
exchange. However, they also gave evidence that it might be
difficult or not possible to find useful multilinear forms within
the realm of algebraic geometry.  For this reason there has existed
a general reluctance among cryptographer to explore multilinear map
constructions even though in some constructions such as the
Boneh-Goh-Nissim~\cite{BGN05} slightly homomorphic encryption
system, or the Boneh-Sahai-Waters~\cite{BSW06} Traitor Tracing
scheme, there appears to exist direct generalizations of bilinear
map solutions.

Very recently,  Garg, Gentry, and Halvei~\cite{GGHslides} announced
a surprising result. Using ideal lattices they produced a candidate
mechanism that would approximate or be the moral equivalent of
multilinear maps for many applications. Speculative applications
include translations of existing bilinear map constructions and
direct generalizations as well as future applications. While the
development and cryptanalysis of their tools is at a nascent stage,
we believe that their result opens an exciting opportunity to study
new constructions using a multilinear map abstraction. The promise
of these results is that such constructions can be brought over to
their framework or a related future one. We believe that building
ABE for circuits is one of the most exciting of these problems due
to the challenges discussed above and that existing bilinear map
constructions do not have a direct generalization.

We construct an ABE system of the Key-Policy variety where
ciphertext descriptors are an $n$-tuple $x$ of boolean variables and
keys are associated with boolean circuits of a max depth $\ell$,
where both $\ell$ and $n$ are polynomially bounded and determined at
the time of system setup. Our main construction exposition is for
circuits that are layered (where gates at depth $j$ get inputs from
gates at depth $j-1$) and monotonic (consisting only of $\AND$ plus
$\OR$ gates). Neither one of these impacts are general result as a
generic circuit can be transformed into a layered one for the same
function with a small amount of overhead. In addition, using
DeMorgan's law one can build a general circuit from a monotone
circuit with negation only appearing at the input wires. We sketch
this in Section~\ref{sec:prelim}. We finally note that  using
universal circuits we can realize ``Ciphertext-Policy'' style ABE
systems for circuits.

Our framework of multi-linear maps is that a party can call a group
generator $\GG(1^\lambda,k)$ to obtain a sequence of groups
$\vec{G}=(\G_1, \dots, \G_k)$ each of large prime\footnote{We stress
that our techniques do not rely on the groups being of prime order;
we only need that certain randomization properties hold in a
statistical sense (which hold perfectly over groups of prime order).
Therefore, our techniques generalize to other algebraic settings.}
order $p> 2^\lambda$ where each comes with a canonical generator $g=
g_1,\ldots,g_k$. Slightly abusing notation, if $i+j \leq k$ we can
compute a bilinear map operation on $g_i^a \in \G_i ,g_j^b \in \G_j$
as $e(g_i^a ,g_j^b)=g_{i+j}^{ab}$. These maps can be seen as
implementing multilinear maps\footnote{ We technically consider the
existence of a set of bilinear maps $\{e_{i,j} : G_i \times G_j
\rightarrow G_{i+j} \; | \; i,j \ge 1; \; i+j \le k \}$, but will
often abuse notation for ease of exposition. }. It is the need to
commit to a certain $k$ value which will require the setup algorithm
of our construction to commit to a maximum depth $\ell=k-1$. We will
prove security under a generalization of the decision BDH assumption
that we call the decision $k$-multilinear assumption. Roughly, it
states that given $g, g^{s}, g^{c_1}, \ldots, g^{c_k}$ it is hard to
distinguish $T=g_k^{s \prod_{j \in [1,k]} c_k}$ from a random
element of $\G_k$.

\paragraph{Our Techniques.}
As discussed there is no apparent generalization of the GPSW methods
for achieving ABE for general circuits. We develop new techniques
with a focus on preventing the backtracking attacks we described
above. Intuitively, we describe our techniques as ``move forward and
shift''; this \emph{replaces and subsumes} the linear interpolation
method of GPSW decryption.  In particular, our schemes do not rely
on any sophisticated linear secret sharing schemes, as was done by
GPSW.

Consider a private key for a given monotonic\footnote{Recall that
assuming that the circuit is monotonic is without loss of
generality.  Our method also applies to general circuits that
involve negations.  See Section~\ref{sec:prelim}.} circuit $f$ with
max depth $\ell$ that works over a group sequence $(\G_1, \dots,
\G_k)$. Each wire $w$ in $f$ is associated by the authority with a
random value $r_w \in \Zp$. A ciphertext for descriptor $x$ will be
associated with randomness $s \in \Zp$. A user should with secret
key for $f$ should be able to decrypt if and only if $f(x)=1$.

The decryption algorithm works by computing $g_{j+1}^{s r_w}$ for
each wire $w$ in the circuit that evaluates to 1 on input $x$. If
the wire is 0, the decryptor should not be able to obtain this
value. Decryption works from the bottom up. For each input wire $w$
at depth 1, we compute $g_{2}^{s r_w}$ using a very similar
mechanism to GPSW.

We now turn our attention to OR gates to illustrate how we prevent
backtracking attacks. Suppose wire $w$ is the output of an $\OR$
gate with input wires $A(w),B(w)$ at depth $j$. Furthermore, suppose
on a given input $x$ the wire $A(w)$ evaluates to true and $B(w)$ to
false so that the decryptor has $g_{j}^{s r_{A(w)}}$, but not
$g_{j}^{s r_{B(w)}}$. The private key components associated with
wire $w$ are:
\[
g^{a_w},  \ g^{b_w},  \ g_j^{r_w - a_w \cdot r_{A(w)} }, \   g_j^{r_w -b_w \cdot r_{B(w) }}
\]
for random $a_w,b_w$. To move decryption onward the algorithm first
computes
$$e\left(g^{a_w}, g_{j}^{s r_{A(w)}}\right) = g_{j+1}^{ s a_w
r_{A(w)}}$$
This is the move forward step. Then it computes
$$e\left(g^s,
g_j^{r_w - a_w \cdot r_{A(w)} }\right)= g_{j+1}^{ s(r_w  -a_w
r_{A(w)} }$$
This is the shift step. Multiplying these together
gives the desired term $g_{j+1}^{s r_w}$.

Let's examine  backtracking attacks in this context. Recall that the
attacker's goal would be to compute $g_{j}^{s r_{B(w)}}$ even though
wire $B(w)$ is 0, and propagate this forward. From the output term
and the fourth key component the attacker can actually inverse the
shift process on the $B$ side and obtain $g_{j+1}^{ s a_w
r_{A(w)}}$, however, since the map $e$ works only in the ``forward''
direction, it is not possible to invert the move forward step and
complete the attack. The crux of our security lies in this idea. In
the main body of this paper we give our formal proof that captures
this intuition.

The $\AND$ gate mechanism has a similar shift and move forward
structure, but requires both inputs for decryption. If this process
is applied iteratively, to an output gate $\tilde{w}$ then one
obtains $g_k^{s r_{\tilde{w}}}$. A final header portion of the key
and decryption mechanism is used to obtain the message. This portion
is similar to prior work.

The details of our scheme and security proof are below.

\section{Preliminaries}
\label{sec:prelim}

In this section, we provide some preliminaries.

\subsection{General Circuits vs. Monotone Circuits}

We begin by observing that there is a folklore transformation that
uses De Morgan's rule to transform any general Boolean circuit into
an equivalent monotone Boolean circuit, with negation gates only
allowed at the inputs.  For completeness, we sketch the construction
here.

Given a Boolean circuit $C$, consider the Boolean circuit
$\tilde{C}$ that computes the negation of $C$.  Note that such a
circuit can be generated by simply recursively applying De Morgan's
rule to each gate of $C$ starting at the output gate.  Note that in
this circuit $\tilde{C}$ each wire computes the negation of the
corresponding wire in $C$.

Now, we can construct a monotone circuit $M$ by combining $C$ and
$\tilde{C}$ as follows: take each negation gate in $C$, eliminate
it, and replace the output of the negation gate by the corresponding
wire in $\tilde{C}$.  Do the same for negation gates in $\tilde{C}$,
using the wires from $C$.  In the end, this will yield a monotone
circuit $M$ with negation gates remaining only at the input level,
as desired.  The size of $M$ will be no more than twice the original
size of $C$, and the depth of $M$ will be identical to the depth of
$C$.  The correctness of this transformation follows trivially from
De Morgan's rule.

As a result, we can focus our attention on monotone circuits.  Note
that inputs to the circuit correspond to attributes, and since we
are in the ``small universe'' setting, we can simply introduce
explicit attributes corresponding to the negation of attributes not
present.

\subsection{Multi-linear maps}

We assume the existence of a group generator $\GG$, which takes as input a
security parameter $n$ and a positive integer $k$ to indicate the number of allowed
pairing operations. $\GG(1^\lambda,k)$ outputs a sequence of groups $\vec{G}=(\G_1, \dots, \G_k)$ each
of large prime order $p>2^\lambda$. In addition,  we let $g_i$ be a canonical generator of
$\G_i$ (and is known from the group's description). We let $g=g_1$.

We assume the existence of a set of bilinear maps $\{e_{i,j} : G_i
\times G_j \rightarrow G_{i+j} \; | \; i,j \ge 1; \; i+j \le k \}$.
The map $e_{i,j}$ satisfies the following relation:

$$e_{i,j}\left( g_i^a, g_j^b \right) = g_{i+j}^{ab} \; : \; \forall
a,b \in \mathbb{Z}_p$$

We observe that one consequence of this is that $e_{i,j} (g_i, g_j) =
g_{i+j}$ for each valid $i,j$.

When the context is obvious, we will sometimes abuse notation drop the
subscripts $i,j$, For example, we may simply write:

$$e\left(g_i^a, g_j^b\right) = g_{i+j}^{ab}$$

We define the decision $k$-multilinear problem as follows.
A challenger runs $\GG(1^\lambda,k)$. Then it picks random $s,c_1,\ldots,c_{k}$.

$g, g^{s}, g^{c_1}, \ldots, g^{c_k}$ it is hard to distinguish $T=g_k^{s \prod_{j \in [1,k]} a_k}$ from
a random group element in $\G_k$.

The decision $k$-multilinear assumption is that no poly-time attacker can win this game with
non-negligible advantage in $\lambda$.

\subsection{Circuit Notation}
\label{sec:circuit-notation}

We now define our notation for circuits that adapts the model and notation of
Bellare, Hoang, and Rogaway~\cite{BHR12} (Section 2.3).
For our application we restrict our consideration to certain classes of boolean circuits.
First, our circuits will have a single output gate. Next, we will consider layered circuits.
In a layered circuit a gate at depth $j$ will receive both of its inputs from wires at depth
$j-1$. Finally, we will restrict ourselves to monotonic circuits where gates are either
AND or OR gates of two inputs.
\footnote{
These restrictions are mostly useful for exposition and do not impact functionality.
General circuits can be built from non-monotonic circuits. In addition, given  a circuit
an equivalent layered exists that is larger by at most a polynomial factor.}

Our circuits will be a five tuple $f=(n,q,A,B,\gatetype)$. We let $n$ be the number of inputs
and $q$ be the number of gates. We define $\inputs = \{1,\ldots,n \}$, $\wires= \{  1,\ldots, n+q \}$,
and $\gates=\{ n+1, \ldots, n+q\}$. The wire $n+q$ is the designated output wire.
$A: \gates \rightarrow \wires / \outputwire$ is a function where $A(w)$ identifies $w$'s first incoming
wire and $B: \gates \rightarrow \wires / \outputwire$ is a function where $B(w)$ identifies $w$'s second
incoming wire. Finally, $\gatetype: \gates \rightarrow \{\AND, \OR \}$ is a function that identifies a gate
as either an AND or OR gate.

We require that $w > B(w) > A(w)$. We also define a function $\depth(w)$ where if $w \in \inputs$
$\depth(w)=1$ and in general $\depth(w)$ of wire $w$ is equal to the shortest path to an input wire
plus 1. Since our circuit is layered we require that for all $w \in \gates$ that if $\depth(w)=j$ then
$\depth(A(w)) = \depth(B(w))= j-1$.

We will abuse notation and let $f(x)$ be the evaluation of the circuit $f$ on
input $x \in \bit^n$. In addition, we let $f_w(x)$ be the value of wire $w$
of the circuit on input $x$.

\section{Our Construction}

We now describe our construction. Our construction is of the
Key-Policy form where a key generation algorithm takes in the
description of a circuit $f$ and encryption takes in an input $x$ and
message $M$. A user with secret key for $f$ can decrypt if and only if
$f(x)=1$. The system is of the ``public index'' variety in that only
the message $M$ is hidden while $x$ can be efficiently discovered from
the ciphertext.

The setup algorithm will take as inputs a maximum depth $\ell$ of all
the circuits as well as the input size $n$ for all ciphertexts. All
circuits $f$ in our system will be of depth $\ell$ (have the
outputgate at depth $\ell$) and be layered as discussed in
Section~\ref{sec:circuit-notation}. Using layered circuits and having
all circuits be of the same depth is primarily for ease of exposition,
as we believe that our construction could directly be adapted to the
general case. The fact that setup defines a maximum depth $\ell$ is
more fundamental as the algorithm defines a $k=\ell+1$ group sequence
a $k$ pairings.


\paragraph{Setup($1^\lambda, n, \ell$)}
The setup algorithm takes as input, a security parameter $\lambda$,  the maximum depth $\ell$ of a circuit, 
and the number of boolean inputs $n$. 

It then runs $\GG(1^\lambda, k =\ell+1)$ and 
of groups $\vec{\G} = (\G_1,\ldots,\G_{k})$ of prime order $p$, with canonical generators $g_1,\ldots,g_k$.
We let $g = g_1$. Next, it chooses random $\alpha \in \Zp$ and $h_1,\ldots,h_\ell \in \G_1$.

The public parameters, $\PP$, consist of the group sequence description plus:
\[
 g_k^{\alpha}, h_1,\ldots, h_{\ell}
\]

The master secret key $\MSK$ is  $(g_{k-1})^{\alpha}$.

\paragraph{Encrypt($\PP, x \in \bit^n, M \in \bit)$}
The encryption algorithm takes in the public parameters, an descriptor input $x \in \bit^n$,
and a  message bit $M \in \bit$. 

The encryption algorithm chooses a random $s$. If $M=0$ it sets $C_M$ to be a random group element in $\G_k$;
otherwise it lets $C_M= (g_k^{\alpha})^s$. Next, let $S$ be the set such of $i$ such that $x_i=1$.

The ciphertext is created as
\[
\CT= ( C_M, \  g^s, \ \forall i \in S \ \  C_i = h_i^{s}  )
\]

\paragraph{KeyGen($\MSK,  f=(n,q,A,B,\gatetype) $)}
The algorithm takes in the master secret key and a description $f$ of a circuit.
Recall, that the circuit has $n+q$ wires with $n$ input wires, $q$ gates and the wire $n+q$
designated as the output wire.

The key generation algorithm chooses random $r_1,\ldots, r_{n+q} \in \Zp$, where we 
think of randomness $r_w$ as being associated with wire $w$.
The algorithm produces a ``header'' component
\[
K_H = (g_{k-1})^{\alpha -r_{n+q} }
\]

Next, the algorithm generates key components for every wire $w$. The structure
of the key components depends upon if $w$ is an input wire, an OR gate, or an AND gate.
We describe how it generates components for each case.

\begin{itemize}

\item \emph{Input wire}\\
By our convention if $w \in [1,n]$ then it corresponds to the $w$-th input.
The key generation algorithm chooses random $z_w \in \Zp$. 

The key components are:
\[
  K_{w,1}= g^{r_w} h_{w}^{z_w}, \  K_{w,2} = g^{- z_w}
\]

\item \emph{OR gate}\\
Suppose that wire $w \in \gates$ and that $\gatetype(w)=\OR$. In addition, let $j=\depth(w)$ be the depth of wire $w$.
The algorithm will choose random
$a_w, b_w \in \Zp$.  
Then the algorithm creates key components:
\[
K_{w,1}= g^{a_w}, \ K_{w,2}=g^{b_w},  \ K_{w,3}=g_j^{r_w - a_w \cdot r_{A(w)} }, \   K_{w,4}=g_j^{r_w -b_w \cdot r_{B(w) }}
\]

\item \emph{AND gate}\\
Suppose that wire $w \in \gates$ and that $\gatetype(w)= \AND $. In addition, let $j=\depth(w)$ be the depth of wire $w$.
The algorithm will choose random
$a_w, b_w \in \Zp$.  
\[
K_{w,1}=g^{a_w}, \ K_{w,2}=g^{b_w}, \  K_{w,3}=g_j^{r_w - a_w \cdot r_{A(w)} - b_w \cdot r_{B(w)} }
\]

\end{itemize}

We will sometimes refer to the $K_{w,3}, K_{w,4}$ of the $\AND$ and $\OR$ gates as the ``shift'' components.
This terminology will take on more meaning when we see how they are used during decryption.

The secret key $\SK$ output consists of the description of $f$, the header component $K_H$ and
the key components for each wire $w$.

\paragraph{Decrypt($\SK,\CT$)}
Suppose that we are evaluating decryption
for a secret key associated with a circuit  $f=(n,q,A,B,\gatetype)$
and a cipherext with input $x$. We will be able to decrypt if $f(x)=1$.

We begin by observing that the goal of decryption should be to 
compute $g_k^{\alpha s}$ such that we can test if this is equal to $C_M$.
The algorithm begins with a header computation and lets
First, there is a header computation where we compute
$
E' = e(K_H), g^s) =  e(g_{k-1}^{\alpha -r_{n+q}},g^s)= g_k^{\alpha s} g_k^{-r_{n+q} \cdot s}
$
Our goal is now reduced to computing $g_k^{r_{n+q} \cdot s}$.

Next, we will evaluate the circuit from the bottom up. Consider wire $w$ at depth $j$; if $f_w(x)=1$ then,
our algorithm will compute $E_w= (g_{j+1})^{s r_w}$. (If $f_w(x)=0$ nothing needs to be computed for that wire.)
Our decryption algorithm proceeds iteratively starting with computing $E_1$ and proceeds in order to finally
compute $E_{n+q}$. Computing these values in order ensures that the computation on
a depth $j-1$ wire (that evaluates to 1) will be defined before computing for a depth $j$ wire.
We show how to compute $E_w$ for all $w$ where $f_w(x)=1$, again breaking the 
cases according to whether the wire is an input, $\AND$ or $\OR$ gate.

\begin{itemize}

\item \emph{Input wire}\\ 
By our convention if $w \in [1,n]$ then it corresponds to the $w$-th input. Suppose that
$x_w = f_w(x)=1$. The algorithm computes:
\[
 E_w= e(K_{w,1} ,  g^s ) \cdot e(K_{w,2} ,  C_w  )     =   e(g^{r_w} h_{w}^{z_w} ,g^s) \cdot  e( g^{ -z_w}, h_{w}^s) = g_{2}^{s r_w}
\]
We observe that this mechanism is similar to many existing ABE schemes. 

\item \emph{OR gate}\\
Consider a wire $w \in \gates$ and that $\gatetype(w)=\OR$. In addition, let $j=\depth(w)$ be the depth of wire $w$.
Suppose that $f_w(x)=1$. If $f_{A(w)}(x)=1$ (the first input evaluated to 1) then we compute:
\[
E_w=   e(E_{A(w)} , K_{w,1})   \cdot      e(  K_{w,3}   ,g^s)
=  e(g_{j}^{s r_{A(w)} },  g^{a_w})  \cdot  e(g_j^{r_w -a_w \cdot r_{A(w) } }, g^s)
=  (g_{j+1})^{s r_w}
\]

Alternatively, if $f_{A(w)}(x)=0$, but $f_{B(w)}(x)=1$, then we compute:
\[
E_w=   e(E_{B(w)} , K_{w,2})   \cdot      e(  K_{w,4}   ,g^s)
=  e(g_{j}^{s r_{B(w)} },  g^{b_w})  \cdot  e(g_j^{r_w -b_w \cdot r_{B(w) } }, g^s)
=  (g_{j+1})^{s r_w}
\]

Let's exam this mechanism for the case where the first input is 1 ($f_{A(w)}(x)=1$).
In this case the algorithm  ``moves'' the value $E_{A(w)}$ from group $\G_{j}$ to
group $\G_{j+1}$ when pairing it with   $K_{w,1}$.  It then multiplies it by $e(  K_{w,3}   ,g^s)$
which ``shifts'' that result to $E_{w}$. 

Suppose that $f_{A(w)}(x)=1$, but $f_{B(w)}(x)=0$. A critical feature of the mechanism is that
an attacker cannot perform a ``backtracking'' attack to compute $E_{B(w)}$. The reason is that
the pairing operation cannot be reverse to go from group $\G_{j+1}$ to group $\G_{j}$. 
If  this were not the case, it would be
debilitating for security as gate $B(w)$ might have fanout greater than 1. This type of backtracking
attacking is why existing ABE constructions are limited to circuits with fanout of 1.

\item \emph{AND gate}\\
Consider a wire $w \in \gates$ and that $\gatetype(w)=\AND$. In addition, let $j=\depth(w)$ be the depth of wire $w$.
Suppose that $f_w(x)=1$. Then  $f_{A(w)}(x)=f_{B(w)}(x)=1$  and we compute:
\[
E_w= e( E_{A(w)}, K_{w,1})   \cdot e(E_{B(w)} , K_{w,2}) \cdot  e(  K_{w,3}   ,g^s) \]
\[
=e(g_{j}^{s r_{A(w)} },  g^{a_w}) \cdot   e(g_{j}^{s r_{B(w)} },  g^{b_w}) \cdot
e(g_j^{r_w-a_w \cdot r_{A(w)} - c_w \cdot r_{B(w) } }  , g^s) 
=  (g_{j+1})^{s r_w} 
\]

\end{itemize}
\vspace{.1 in}§

If the $f(x)=f_{n+q}(x)=1$, then  the algorithm will compute $E_{n+q}=g_k^{r_{n+q} \cdot s}$.
It finally computes $E' \cdot E_{n+q}=g_k^{\alpha s}$ and tests if this equals $C_M$, outputting $M=1$ if so
and $M=0$ otherwise. Correctness holds with high probability.

\paragraph{A Few Remarks}
We  end this section with a few remarks. First, the encryption algorithm takes as input a single
bit message. In this setting we could imagine encoding a longer message by
XORing it with the hash of $g_k^{\alpha s}$. However, we used bit encryption with a testability 
function to better match the lattice translation of the next section.

Our $\OR$ and $\AND$ key components respectively have one and two ``shift'' components.
It is conceivable to have a construction with one shift component for the OR and none for the AND.
However, we designed it this way since it made the exposition of our proof (in particular the distribution
of private keys) easier.

Finally, our construction uses a layered circuit, where a wire at depth $j$ gets its inputs from 
depth $j'=j-1$. We could imagine a small modification to our construction which allowed
$j'$ to be of any depth less than $j$. Suppose this were the case for the first input. Then
instead of $K_{w,1}= g_1^{a_w}$ we might more generally let $K_{w,1}= (g_{j-j'})^{a_w}$.
However, we stick to describing and proving the layered case for simplicity.

\section{Proof of Security}

We prove (selective) security in the security model given by
GPSW~\cite{GPSW06}, where the input access structures are monotonic
circuits. For a circuit of max size $k-1$ we prove security under the
decision $k$-multilinear assumption.

We show that if there exist a poly-time attacker $\AlgA$ on our ABE
system for circuits of depth $\ell$ and inputs of length $n$ in the
selective security game then we can construct a poly-time algorithm on
the decision $\ell+1$-multilinear assumption with non-neglgibile
advantage. We describe how $\AlgB$ interacts with $\AlgA$.

\paragraph{Init}
$\AlgB$ first receives the $\ell+1$-multilinear problem where it is
given the group description $\vec{\G} = (\G_1,\ldots,\G_{k})$ and an
problem instance $g, g^{s}, g^{c_1}, \ldots, g^{c_k}, T$.  $T$
is either $g_k^{s \prod_{j \in [1,k]} a_j}$ or a random group element
in $\G_k$. (Note we slightly changed the variable names in the problem
instance to better suit our proof.)

Next, the attacker declares the challenge input $x^* \in \bit^n$.

\paragraph{Setup}
$\AlgB$ chooses random $y_1,\ldots, y_n  \in \Zp$.
For $i \in [1,n]$ set
\[
h_i = 
\begin{cases}
g^{y_i} & \mathrm{if~} x^*_i=1 \\
g^{y_i + c_1} & \mathrm{if~} x^*_i=0
\end{cases}
\]

\noindent \textbf{Remark.}  Note that over $\mathbb{Z}_p$, the above
choices of $h_i$ are distributed identically with the ``real life''
distribution.  More generally, what we need is that $g^{y_i}$ is
statistically close to, or indistinguishable from, $g^{y_i+c_1}$.

Next, $\AlgB$ sets $g_k^{\alpha}$= $g_k^{ \xi + \prod_{i \in [1,k]}
  c_i}$, where $\xi$ is chosen randomly. It computes this using
$g^{c_1}, \ldots, g^{c_k}$ from the assumption, by means of the
iterated use of the pairing function.

\noindent \textbf{Remark.}  Here we need that $g_k^{ \xi + \prod_{i
    \in [1,k]} c_i}$ is statistically close to, or indistinguishable
from, $g_k^{ \xi}$.  This holds perfectly over $\mathbb{Z}_p$.

\paragraph{Challenge Ciphertext} 
Let $S^* \subseteq [1,n]$ be the set of input indices  where $x^*_i=1$.
$\AlgB$ creates the challenge ciphertext
as:
\[
\CT= ( T, \  g^s, \ \forall j \in S^* \ C_i = (g^{s})^{y_j}    )
\]

If $T=g_k^{s \prod_{j \in [1,k]} ac_k}$ then this is an encryption of 1; otherwise if $T$ was chosen random
in $\G_k$ then w.h.p. it is an encryption of 0.

\paragraph{KeyGen Phase} 
Both key generation phases are executed in the same manner by the
reduction algorithm.  Therefore, we describe them once here. The
attacker will give a circuit $ f=(n,q,A,B,\gatetype)$ to the reduction
algorithm such that $f(x^*)=0$.

We can think of the proof as having some invariant properties on the
depth of the gate we are looking at. Consider a gate $w$ at depth $j$
and the simulators viewpoint (symbolically) of $r_w$.  If
$f_w(x^*)=0$, then the simulator will view $r_w$ as the term $c_1
\cdot c_2 \cdots c_{j+1}$ plus some additional known randomization
terms. If $f_w(x^*)=1$, then the simulator will view $r_w$ as the 0
plus some additional known randomization terms.  If we can keep this property
intact for simulating the keys up the circuit, the simulator will view
$r_{n+q}$ as $c_1 \cdot c_2 \cdots c_\ell$ . This will allow for it to
simulate the header component $K_H$ by cancellation.

We describe how to create the key components for each wire $w$. Again,
we organize key component creation into input wires, $\OR$ gates, and $\AND$ gates.

\begin{itemize}

\item \emph{Input wire}\\
Suppose $w \in [1,n]$ and is therefore by convention an input wire. 

If $(x^*)_{w}=1$ then we choose $r_w$ and $z_w$ at random
(as is done honestly).
The key components are:
\[ 
(K_{w,1}=g^{r_w}h_{w}^{z_w}, \ 
K_{w,2} = g^{z_w})
\]

If  $(x^*)_{w}=0$ then we let $r_w=c_1 c_2 + \eta_i$ and $z_w=-c_2 + \nu_i$,
where $\eta_i$ and $\nu_i$ are randomly chosen elements. 
The key components are:
\[ 
(K_{w,1}=g^{c_1 c_2 + \eta_w}h_{w}^{-c_2+\nu_w}, \ K_{w,2}= g^{-c_2+\nu_w} ) 
=   (g^{-c_2 y_w + \eta_w + (y_w+c_1)\nu_w}  , g^{-c_2+\nu_w}    )
\]
Note a cancellation occurred that allowed for the first term to be computed. 
Observe that in  both of these values are simulated consistent with our invariant.

\noindent \textbf{Remark.}  Here we need that $g^{-c_2 y_w + \eta_w +
  (y_w+c_1)\nu_w}$ is appropriately close to a randomly chosen
element.  This holds perfectly over $\mathbb{Z}_p$.

\item \emph{OR gate}\\
Now we consider a wire $w \in \gates$ and that $\gatetype(w)=\OR$. 
In addition, let $j=\depth(w)$ be the depth of wire $w$.
If $f_w(x^*) =1$, then we simply set $a_w, b_w, r_w$ at random to values
chosen by $\AlgB$.
Then the algorithm creates key components:
\[
K_{w,1}= g^{a_w}, \  K_{w,2}=g^{b_w},  \  K_{w,3} = g_j^{r_w-a_w \cdot r_{A(w)} }, \   K_{w,4} = g_j^{r_w-b_w \cdot r_{B(w) }}
\]

If $f_w(x^*) =0$, then we set $a_w= c_{j+1}+\psi_w$ and $b_w=c_{j+1} + \phi_w$ and 
$r_w= c_1\cdot c_2 \cdots c_{j+1} + \eta_w$,
where $\psi_w, \phi_w, \eta_w$ are chosen randomly.
Then the algorithm creates key components:
\[
K_{w,1}= g^{c_{j+1}+\psi_w}, \  
K_{w,2}= g^{c_{j+1}+\psi_w},  
\]
\[
K_{w,3}= g_j^{\eta_w - c_{j+1}\eta_{A(w)} - \psi_w (c_1 \cdots c_{j} + \eta_{A(w)})},
\   
K_{w,4}= g_j^{\eta_w -  c_{j+1}\eta_{B(w)} - \phi_w (c_1 \cdots c_{j} + \eta_{B(w)})}
\]
$\AlgB$ is able to create the last two key components due to a cancellation. Since both the
$A(w)$ and $B(w)$ gates evaluated to 0 we had $r_{A(w)} = c_1 \cdots c_{j} + \eta_{A(w)}$ and
similarly for $r_{B(w)}$.  Note that computing $g_j^{c_1 \cdots c_j}$ is possible using the multi-linear
maps.

\noindent \textbf{Remark.}
Here we need that $g_j^{\eta_w - \psi_w (c_1 \cdots c_{j} )}$ is appropriately close to a randomly chosen element
(the given terms dominate the others).
This holds perfectly over $\mathbb{Z}_p$.

\item \emph{AND gate}\\
Now we consider a wire $w \in \gates$ and that $\gatetype(w)=\OR$. 
In addition, let $j=\depth(w)$ be the depth of wire $w$.

If $f_w(x^*) =1$,  then we simply set $a_w, b_w, r_w$ at random to values
known by $\AlgB$.
Then the algorithm creates key components:
\[
K_{w,1}=g^{a_w} , \ K_{w,2}=g^{b_w}, \  K_{w,3}=g_j^{r_w - a_w \cdot r_{A(w)} - b_w \cdot r_{B(w)} }
\]

If $f_w(x^*) =0$ and $f_{A(w)}(x^*)=0$, 
then $\AlgB$ sets  $a_w = c_{j+1} + \psi_w, b_w = \phi_w$ and $r_w= c_1\cdot c_2 \cdots c_{j+1} + \eta_w$,
where $\psi_w, \phi_w, \eta_w$ are chosen randomly.
Then the algorithm creates key components:
\[
K_{w,1}= g^{c_{j+1} + \psi_w} , \ 
K_{w,2}=g^{\phi_w}, \  
K_{w,3}=   g_j^{\eta_w - \psi_w c_1 \cdots c_j - (c_{j+1} + \psi_w) \eta_{A(w)} - \phi_w (r_{B(w)})}
\]
$\AlgB$ can create the last component due to cancellation. Since the
$A(w)$ gate evaluated to 0, we have $r_{A(w)} = c_1\cdot c_2 \cdots c_{j} + \eta_{A(w)}$.
Note that $g_j^{r_{B(w)}}$ is always computable regardless of whether $f_{A(w)}(x^*)$ evaluated
to 0 or 1, since $g_j^{c_1 \cdots c_j}$ is always computable using the multilinear maps.

The case where $f_{B(w)}(x^*)=0$ and $f_{A(w)}(x^*)=1$ is performed  in a symmetric to what is above, with the
roles of $a_w$ and $b_w$ reversed.

\noindent \textbf{Remark.}
Here we need that $g_j^{\eta_w - (\psi_w + \phi_w) \cdot (c_1 \cdots c_{j} )}$ is appropriately close to a randomly chosen element
(the given terms dominate the others).
This holds perfectly over $\mathbb{Z}_p$.

\end{itemize}

For the output gate we chose $\eta_w$ at random.
Thus, at the end we have $r_{n+q} =\prod_{i \in [1,k]} c_i + \eta_{n+q}$ for the output gate. 
This gives us a final cancellation in computing the ``header'' component of the key as
 $K_H= (g_{k-1})^{\alpha -r_{n+q}} = (g_{k-1})^{\xi-\eta_w}$.

\paragraph{Guess}
$\AlgB$ receives back the guess $M' \in \bit$ of the message from $\AlgA$. If $M'=1$ it guesses
that $T$ is a tuple; otherwise, it guesses that it is random.

\bibliographystyle{alpha}
\bibliography{functional}

\end{document}